\documentclass[aps,prl,reprint,letterpaper,amsfonts,amssymb,amsmath,showkeys,showpacs,final]{revtex4-1}
\pdfoutput=1

\usepackage[utf8]{inputenc}
\usepackage[T1]{fontenc}
\usepackage[english]{babel}

\usepackage{mathrsfs}

\DeclareFontFamily{U}{rsfs}{}         
\DeclareFontShape{U}{rsfs}{m}{n}{<5> rsfs5 <6><7> rsfs7          %
  <8><9><10><10.95><12><14.4><17.28><20.74><24.88> rsfs10}{}     %
\DeclareMathAlphabet{\mathfs}{U}{rsfs}{m}{n} 

\usepackage[autostyle,english=american]{csquotes}

\usepackage{microtype}

\begin{document}
\title{Interpreting the isolated horizon boundary condition in terms of higher gauge theory}
\author{Thomas Zilker}
\email{Thomas.Zilker@gravity.fau.de}
\affiliation{Institute for Quantum Gravity\\ Friedrich-Alexander-Universität Erlangen-Nürnberg\\ Staudtstraße 7/B2, 91056 Erlangen, Germany}

\date{\today}

\begin{abstract}
The purpose of this letter is to point out a relation between the boundary condition satisfied by spherically symmetric isolated horizons (formulated in terms of Ashtekar-Barbero variables) and the source-target matching condition (also known as fake flatness condition) in higher gauge theory. This relation may prove useful in the attempt to quantize the isolated horizon boundary condition which is in turn a possible starting point for the search for black hole solutions in the full theory of loop quantum gravity. Also, since a 2-connection is the mathematical object required for describing the parallel transport of 1-dimensional objects, the relation presented in this letter may provide further insight into the coupling of LQG to string-like objects investigated in other contexts.\\
\end{abstract}

\pacs{04.60.Pp, 04.70.-s}
\keywords{black holes, loop quantum gravity, isolated horizons, higher gauge theory, 2-connections}

\maketitle

\section{The isolated horizon boundary condition}
\label{sec:IHBC}

The description of black holes is an important test for any theory of quantum gravity. Loop quantum gravity (LQG) descriptions of black holes make use of the concept of isolated horizons first introduced in \citep{Ashtekar1999}. In these works black holes are modeled as isolated horizons which are also inner boundaries of the spacetime manifold. The presence of boundaries gives rise to additional terms in the action as well as in the symplectic structure. The boundary term in the latter can be given the form of an $SU(2)$ Chern-Simons (CS) theory with the pullback of the Ashtekar connection to (a spatial slice of) the isolated horizon as the CS connection. Since the Ashtekar connection is continuous in the classical theory, its pullback to the horizon is completely determined by its values in the bulk. Furthermore, on isolated horizons the pullbacks of the elementary phase space variables ($\underset{\Leftarrow}{A}$,$\underset{\Leftarrow}{\Sigma}$) need to satisfy a specific boundary condition. In the case of spherically symmetric isolated horizons, this condition reads
\begin{equation}
F(\underset{\Leftarrow}{A})^{i} = c \, \underset{\Leftarrow}{\Sigma}^{i} \, ,
\label{eqn:IHBC}
\end{equation}
where c is a constant related to the surface area of the horizon. In the derivation of loop quantum gravity from general relativity, the configuration space is extended to the space of generalized connections. In contrast to ordinary connections, these need not be continuous anymore. Thus, in the quantum theory, the boundary degrees of freedom (encoded in the CS connection $\underset{\Leftarrow}{A}$) are independent of those bulk degrees of freedom that are encoded in the Ashtekar connection $A$. Nevertheless, there is still equation \eqref{eqn:IHBC} -- or, more precisely, a quantum operator version thereof -- linking $\underset{\Leftarrow}{A}$ with some of the bulk degrees of freedom (namely those encoded in $\underset{\Leftarrow}{\Sigma}$). LQG treatments of black holes usually proceed by first quantizing bulk and boundary degrees of freedom separately. For the bulk part, this leads to the standard LQG Hilbert space spanned by spin networks. The boundary degrees of freedom, on the other hand, are described by a CS theory with defects, where the defects are located at those points at which the boundary is punctured by a spin network. Such defects carry spin labels and in principle these are new, independent quantum numbers. The full Hilbert space of the system is then given by the tensor product of these two Hilbert spaces. The role of the isolated horizon boundary condition is now to bring the two parts together by imposing a quantum version of equation \eqref{eqn:IHBC}. This quantum equation can be thought of as
\begin{equation}
\operatorname{id}_{\mathcal{H}_{\text{LQG}}} \otimes \widehat{F(\underset{\Leftarrow}{A})}^{i} = c \, \widehat{\underset{\Leftarrow}{\Sigma}}^{i} \otimes \operatorname{id}_{\mathcal{H}_{CS}} \, .
\end{equation}

Effectively this leads to the CS defects being labeled by the same spins as the edges of the spin network piercing the horizon in the corresponding punctures. Restricting the boundary Hilbert space in this way, the isolated horizon boundary condition thus has a crucial impact on calculations of black hole entropy \citep{Smolin1995,Ashtekar2000,Domagala2004,Kaul1998,Kaul2000,Engle2009,Engle2010}. The main purpose of this paper is to introduce a new perspective on the isolated horizon boundary condition by regarding it in terms of higher gauge theory.

\section{Higher gauge theory: 2-groups, 2-connections and 2-holonomies}
\label{sec:HGT}

As we want to give an interpretation of the isolated horizon boundary condition in terms of higher gauge theory, it is necessary to give a brief introduction to higher gauge theory first. In doing so, we restrict ourselves to topics that are needed in order to follow the rest of the article. A more thorough exposition to higher gauge theory can be found in \citep{Baez2011} and references therein.\\
Higher gauge theory is a generalization of ordinary gauge theory. In standard gauge theory the dynamic variable is a connection on a principle G-bundle over the spacetime manifold. By constructing holonomies one can use this connection to associate group elements to paths in the spacetime manifold. The aim of higher gauge theory is to extend this construction to higher dimensional objects. At the next level, one wants to associate group elements not only to paths but also to homotopies (\enquote{paths between paths}) in a consistent way. The mathematical structure required for this is a 2-connection on a principle $\mathcal{G}$-2-bundle, where $\mathcal{G}$ is a (strict) 2-group.\\
A 2-group is a generalization of a group in the sense of category theory. Thinking of a group as a category with a single object and with all morphisms being isomorphisms, it is straight forward to define a (strict) 2-group as a (strict) 2-category with a single object and in which every 1-morphism and every 2-morphism is an isomorphism. While this concept of a 2-group is rather abstract, there is a theorem (see e.g. section 3.3 of \citep{Baez2011}) that makes it possible to identify strict 2-groups with crossed modules of groups, i.e. with quadruples ($G$,$H$,$t$,$\alpha$). Here $G$ and $H$ denote groups, $t$ is a group homomorphism $t: H \rightarrow G$ and $\alpha$ is a group homomorphism $\alpha: G \rightarrow \operatorname{Aut}(H)$, that is $\alpha$ gives a group action of G on H. The homomorphisms $t$ and $\alpha$ are subject to the two conditions
\begin{eqnarray}
t(\alpha(g)h) &= g t(h) g^{-1}\\
\alpha(t(h))h' &= h h' h^{-1}
\end{eqnarray}
for all $g \in G$ and $h,h' \in H$. A strict 2-group $\mathcal{G}$ encoded in a crossed module ($G$,$H$,$t$,$\alpha$) is called a Lie 2-group, if $G$ and $H$ are Lie groups and if $t$ and $\alpha$ are Lie group homomorphisms. Just as any Lie group has a corresponding Lie algebra one can associate a Lie 2-algebra to every Lie 2-group. The Lie 2-algebra corresponding to ($G$,$H$,$t$,$\alpha$) is encoded in the differential crossed module ($\mathfrak{g}$,$\mathfrak{h}$,$\underline{t}$,$\underline{\alpha}$), where $\mathfrak{g}$ and $\mathfrak{h}$ denote the Lie algebras corresponding to the Lie groups $G$ and $H$ and $\underline{t}$ and $\underline{\alpha}$ are the Lie algebra homomorphisms induced by $t$ and $\alpha$.\\
With this notation settled, a 2-connection is now given (at least on the trivial principal $\mathcal{G}$-2-bundle) by a pair ($A$,$B$) of a $\mathfrak{g}$-valued 1-form $A$ and an $\mathfrak{h}$-valued 2-form $B$ subject to the restriction
\begin{equation}
F(A) = \underline{t}(B) \, .
\label{eqn:STMC}
\end{equation}
This condition is necessary to ensure that the 2-holonomy 2-functor associated to the 2-connection ($A$,$B$) actually takes values in the Lie 2-group $\mathcal{G}$. Just as an ordinary holonomy, the 2-holonomy 2-functor associates elements of $G$ to paths in the base manifold of the 2-bundle. Additionally, however, it also associates elements of $H$ to surfaces $S$ (which are the surfaces swiped out by homotopies, i.e. they have to be compact and simply connected) via the construction of surface holonomy. In formulas, the surface holonomy $W_{S}$ associated to the surface $S$ is obtained via
\begin{equation}
W_{S}[A,B] = \mathcal{S} \operatorname{exp} \iint \limits_{S} \hat{\alpha}(h_{x_0 \rightarrow x}[A]) (B) \operatorname{d}^{2} x \, ,
\label{eqn:surface_holonomy}
\end{equation}
where $\mathcal{S} \operatorname{exp}$ denotes the surface ordered exponential as defined in \citep{Aref'eva1980} or \cite{Martins2011} and $\hat{\alpha}: G \rightarrow aut(\mathfrak{h})$ is the map that associates to every element $g \in G$ the element in $aut(\mathfrak{h})$, which is obtained by deriving the group automorphism $\alpha(g): H \rightarrow H$.

\section{LQG variables on spherically symmetric isolated horizons form a 2-connection}

Looking back at formulas \eqref{eqn:IHBC} and \eqref{eqn:STMC}, their similarity is obvious. Interpreting \eqref{eqn:IHBC} as an example of \eqref{eqn:STMC}, we realize that the pair ($\underset{\Leftarrow}{A}$,$c \, \underset{\Leftarrow}{\Sigma}$) actually constitutes a 2-connection on the isolated horizon with respect to the gauge 2-group encoded by the crossed module ($SU(2)$,$SU(2)$,$\operatorname{id}_{SU(2)}$,$\alpha$), with $\alpha$ denoting the action of $SU(2)$ on itself by conjugation.
Now, while every connection $A$ trivially defines a 2-connection via ($A$,$F(A)$), the remarkable part of our discovery lies in the fact that on spherically symmetric isolated horizons, the 2-connection is actually given by the pullbacks ($\underset{\Leftarrow}{A}$,$c\underset{\Leftarrow}{\Sigma}$) of the elementary phase space variables of LQG (up to a constant c). This suggests that the theory govering the horizon degrees of freedom may equivalently be expressed as a higher gauge theory.\\
Rephrasing our findings we come to the conclusion that the search for surfaces solving the isolated horizon boundary condition is equivalent to the search for surfaces, on which the elementary variables of LQG form a 2-connection! A more detailed analysis of the implications of this discovery will be given in the next section.\\
Note also that the gauge 2-group $\mathcal{G}$ involved in the description of the phase space in terms of a 2-connection variable is essentially unique; namely, it is unique if one requires the two Lie groups $G$ and $H$ encoding the gauge 2-group $\mathcal{G}$ to be simply connected. This is due to the fact that to every differential crossed module one can only find one corresponding Lie crossed module, such that its two Lie groups are both simply connected \citep{Martins2011,Knapp2002}.

\section{Implications}

\subsection{Black holes in full LQG}
It was proposed in \citep{Sahlmann2011} that a full LQG description of (spherically symmetric) black holes might be obtained by looking for solutions of a quantum version of the isolated horizon boundary condition. Pursuing this approach one encounters the problem, that there is no well-defined operator associated to $F(A)$ in loop quantum gravity. However, making use of our newly found relation to higher gauge theory, we can equivalently quantize the integrated version of equation \eqref{eqn:IHBC}, which reads
\begin{equation}
h_{\partial S} = W_{S}[\underset{\Leftarrow}{A},\underset{\Leftarrow}{F(A)}] = W_{S}[\underset{\Leftarrow}{A},c\underset{\Leftarrow}{\Sigma}] \, ,
\label{eqn:integratedIHBC}
\end{equation}
where the surface holonomy $W_{S}$ is defined as in equation \eqref{eqn:surface_holonomy} and the first equality in \eqref{eqn:integratedIHBC} is an application of the non-Abelian Stokes' theorem \citep{Aref'eva1980}. Note, however, that in order for a surface H to satisfy the isolated horizon boundary condition, the integrated condition \eqref{eqn:integratedIHBC} has to be satisfied for all simply connected surfaces $S \subset H$. Now this integrated version has been considered in the literature before, but while before it was only clear that any surface satisfying \eqref{eqn:IHBC} also has to satisfy \eqref{eqn:integratedIHBC}, the relation to higher gauge theory allows us to infer the equivalence of these two conditions. The reason is that in higher gauge theory equation \eqref{eqn:STMC} is just the differential version of the condition 
\begin{equation}
t(W_{S}[A,B]) = h_{\partial S}[A] \, ,
\label{eqn:STMC_intVersion}
\end{equation} 
which is a necessary condition for the image of the holonomy 2-functor associated with the pair ($A$,$B$) to be a 2-group. In this sense, surface holonomies are only mathematically well-defined if they are constructed from a 2-connection.\\
This now allows us to quantize equation \eqref{eqn:integratedIHBC} instead of the isolated horizon boundary condition. The advantage of this approach is that there exists a well-defined quantum operator associated to $h_{\partial S}$ in LQG. However, a new problem arises in the task of quantizing the surface holonomies $W_{S}[\underset{\Leftarrow}{A},c\underset{\Leftarrow}{\Sigma}]$ in the LQG setting. Building on a proposal for the quantization of the trace of this operator in \citep{Sahlmann2012}, a first attempt to quantize the full surface holonomies on an isolated horizon has been offered in \cite{Sahlmann2015}. A follow up paper deepening these results is already in progress.

\subsection{2-connections on generic isolated horizons}

Generic static isolated horizons can be described in a similar fashion to spherically symmetric ones. The difference is that their description \citep{Perez2011} requires a pair of connections on the boundary governed by two distinct CS theories (with related coupling constants). Both of these connections satisfy boundary conditions similar to equation \eqref{eqn:IHBC}. More precisely, these boundary conditions read

\begin{eqnarray}
F(\underset{\Leftarrow}{A_{\sigma}})^{i} &= \left( \Psi_{2} + \frac{\sigma^{2} + c}{2} \right) \underset{\Leftarrow}{\Sigma^{i}} =: c_{\sigma} \, \underset{\Leftarrow}{\Sigma^{i}} \, ,\\
F(\underset{\Leftarrow}{A_{\gamma}})^{i} &= \left( \Psi_{2} + \frac{\gamma^{2} + c}{2} \right) \underset{\Leftarrow}{\Sigma^{i}} =: c_{\gamma} \, \underset{\Leftarrow}{\Sigma^{i}} \, ,
\end{eqnarray}

where $\sigma$ and $\gamma$ are real parameters, $\Psi_{2}$ is one of the five complex Weyl scalars used in the Newman-Penrose formalism and $c$ is some curvature scalar definied in \citep{Perez2011}. One major difference to equation \eqref{eqn:IHBC} is that since $\Psi_{2}$ and $c$ are functions on the horizon, the factor of proportionality in these equations is not a constant anymore. The other is, that the two connections $A_{\sigma}$, $A_{\gamma}$ are not determined by the bulk connection. Instead, they are defined via $A_{\sigma} = \Gamma^{i} + \sigma e^{i}$ and $A_{\gamma} = \Gamma^{i} + \gamma e^{i}$, respectively, with $\Gamma^{i}$ denoting the spin connection and $e^{i}$ a spatial triad. So while the form of these boundary conditions is close enough to equation \eqref{eqn:IHBC} such that the higher gauge theory interpretation carries over for ($\underset{\Leftarrow}{A_{\sigma}}$,$c_{\sigma} \, \underset{\Leftarrow}{\Sigma}$) and ($\underset{\Leftarrow}{A_{\sigma}}$,$c_{\gamma} \, \underset{\Leftarrow}{\Sigma}$) separately, these 2-connections no longer have a direct interpretation as phase space variables. 

When we pass to rotating (i.e. non-static) isolated horizons a similar higher gauge theory interpretation is even less obvious. In \citep{Frodden2014} a description of rotating black holes is proposed using the framework described above. In this model the boundary condition at the horizon reads
\begin{equation}
\frac{k}{4\pi} F(\underset{\Leftarrow}{\mathfs{A}})^{i} = \frac{1}{8\pi \beta} \underset{\Leftarrow}{\Sigma}^{i} + p \, \delta^{i}_{1} \, \delta_{N} + p \, \delta^{i}_{1} \, \delta_{S}
\end{equation}
where $\beta$ is the Barbero-Immirzi parameter, $p=\frac{J}{2} + \frac{k}{2}$ (with $J$ the angular momentum of the rotating black hole and $k$ the level of the Chern-Simons theory) and $\delta_{N}$ and $\delta_{S}$ denote delta distributions with support at the north and south pole, respectively. Considering this boundary condition, a higher gauge theory interpretation seems possible if one modifies either $\underset{\Leftarrow}{\mathfs{A}}$ or $\underset{\Leftarrow}{\Sigma}^{i}$ to account for the singularities at the poles. However, note that although the connection $\underset{\Leftarrow}{\mathfs{A}}$ is again a Chern-Simons connection, it is no longer given by the pullback of the (bulk) Ashtekar connection to the horizon (just as in the static, non-spherically symmtric case). Thus the interpretation of solutions to the boundary condition as surfaces on which the elementary phase space variables form a 2-connection does not carry over to these more general situations.

\subsection{String theory on boundaries of spacetime?}

Finally, let us also mention a recent paper by Freidel et al. \citep{Freidel2016}, in which they found that the description of degrees of freedom on boundaries of spacetime is related to descriptions of string matter. Our findings may provide additional evidence for this relation since 2-connections are precisely the mathematical structure required to write down a gauge theory incorporating  1-dimensional objects (strings). The reason is that a 2-connection, via the concept of 2-holonomy, allows to ascribe group elements not only to 1-dimensional objects (world lines of point particles) but also to 2-dimensional objects (world sheets of strings) thus generalizing the concept of parallel transport. However, since the result of \citep{Freidel2016} holds for arbitrary (spherical) boundaries of spacetime, it would be interesting to know, whether the degrees of freedom on every such boundary of spacetime can be encoded in a 2-connection. It is therefore interesting that a boundary condition similar to equation \eqref{eqn:IHBC} enters their derivation. Namely, they require the curvature of the connection to satisfy
\begin{equation}
F(A)^{i}(x) = 2\pi \sum_{p} K^{i}_{p} \delta^{(2)} (x,x_{p}) \, ,
\end{equation}
on the boundary, where the sum is over all spin network punctures p and the $K^{i}_{p}$ denote $SU(2)$ Lie algebra elements. Further insight into the role of 2-connections in loop quantum gravity might thus also improve our understanding of the appearance of string structures in loop quantum gravity. Note that in this paper we only discussed a 2-connection on a spatial slice of (spherically symmetric) isolated horizons. Thus, a natural first step in this direction would be to determine an extension of this 2-connection to the entire, three-dimensional, isolated horizon.

\section{Acknowledgments}

\begin{acknowledgments}
I want to thank Hanno Sahlmann and Derek Wise for helpful discussions. Part of this work was accomplished while I was funded by the Elite Graduate Program of the State of Bavaria.
\end{acknowledgments}

\bibliography{bibliography_IHBC&HGT}

\end{document}